\documentclass[aps,twocolumn,superscriptaddress]{revtex4}
\usepackage{amsmath}
\usepackage{amssymb}
\usepackage{amsthm}
\usepackage{amsfonts}
\usepackage{listings}
\usepackage{enumerate}
\usepackage{latexsym}
\usepackage{color}
\usepackage{bm}
\usepackage{hyperref}
\hypersetup{
 pdfnewwindow=true, colorlinks=true,
 linkcolor=blue, anchorcolor=blue,
 citecolor=blue, filecolor=blue,
 menucolor=blue, urlcolor=blue}

\usepackage{psfrag}

\usepackage{float}
\usepackage{bm}
\usepackage{graphicx}
\usepackage{subfigure}
\usepackage{epstopdf}
\usepackage{tikz-cd}
\RequirePackage[normalem]{ulem} %DIF PREAMBLE
\RequirePackage{color}\definecolor{RED}{rgb}{1,0,0}\definecolor{BLUE}{rgb}{0,0,1} %DIF PREAMBLE
\newcommand{\beq}{\begin{equation}}
\newcommand{\eneq}{\end{equation}}
\begin{document}

\def\pdpbx{Pd$_{3}$Pb$_2$X$_2$}

\def\ie{{\it i.e.},\ }
\def\eg{{\it e.g.}\ }
\def\ea{{\it et al.}}
\input{epsf}

\tolerance 10000

\draft

%\title{High-throughput screening for Weyl Semimetals with S4 Symmetry}
 
\title{Tunable Dirac Semimetals with Higher-order Fermi Arcs in Kagome Lattices \pdpbx~($\text{X}=\text{S}$, $\text{Se}$)}

%in English titles articles and words like to, on, at etc are always spelled with small letters

\author{Simin Nie}
\thanks{These authors contributed equally}
%\email{smnie@stanford.edu}
\affiliation{Department of Materials Science and Engineering, Stanford University, Stanford, California 94305, USA}

\author{Jia Chen}
\thanks{These authors contributed equally}
\affiliation{Zhejiang Lab, Hangzhou, Zhejiang 311121, China}

\author{Changming Yue}
\affiliation{Department of Physics, University of Fribourg, 1700 Fribourg, Switzerland}

\author{Congcong Le}
\affiliation{RIKEN Interdisciplinary Theoretical and Mathematical Sciences (iTHEMS), Wako, Saitama 351-0198, Japan}

\author{Danwen Yuan}
\affiliation{Fujian Provincial Key Laboratory of Quantum Manipulation and New Energy Materials, College of Physics and Energy, Fujian Normal University, Fuzhou 350117, China}
\affiliation{Fujian Provincial Collaborative Innovation Center for Advanced High-field Superconducting Materials and Engineering, Fuzhou 350117, China}

\author{Zhijun Wang}
\affiliation{Beijing National Laboratory for Condensed Matter Physics, Institute of Physics, Chinese Academy of Sciences, Beijing 100190, China}
\affiliation{School of Physics, University of Chinese Academy of Sciences, Beijing 100049, China}

\author{Wei Zhang}
\email{zhangw721@163.com}
\affiliation{Fujian Provincial Key Laboratory of Quantum Manipulation and New Energy Materials, College of Physics and Energy, Fujian Normal University, Fuzhou 350117, China}
\affiliation{Fujian Provincial Collaborative Innovation Center for Advanced High-field Superconducting Materials and Engineering, Fuzhou 350117, China}

\author{Hongming Weng}
\email{hmweng@iphy.ac.cn}
\affiliation{Beijing National Laboratory for Condensed Matter Physics, Institute of Physics, Chinese Academy of Sciences, Beijing 100190, China}
\affiliation{School of Physics, University of Chinese Academy of Sciences, Beijing 100049, China}
\affiliation{Songshan Lake Materials Laboratory, Dongguan, Guangdong 523808, China}

\begin{abstract}
Bulk-boundary correspondence has achieved a great success in the identification of topological states. However, this elegant strategy doesn't apply to the Dirac semimetals (DSMs). Here, we propose that kagome lattices $\text{Pd}_{3}$Pb$_{2}$X$_{2}$ ($\text{X}=\text{S}$, $\text{Se}$)
are unique type-I DSMs without surface Fermi arc states, which are different from the previous well-known DSMs Na$_3$Bi and Cd$_3$As$_2$. $\text{Pd}_{3}$Pb$_{2}$X$_{2}$ are characterized by nontrivial topological invariant $\mathbb{Z}_3$, 
guaranteeing a higher-order bulk-hinge correspondence and the existence of higher-order Fermi arcs, as well as fractional corner charges on the hinges. The type-I DSMs are located at the phase boundaries of several topological phases, including type-II DSMs and three-dimensional weak topological insulators. The phase transitions can be easily manipulated by external strain. Our results provide feasible platforms for the study of these unique DSMs and the related phase transitions. 

%\newline
%\textbf{Keywords:} Dirac semimetals, Bulk-hinge correspondence, High-order Fermi arcs, Fractional corner charges, Kagome lattices

\end{abstract}

\maketitle
%\section{Introduction}

Over the decade, Dirac semimetals (DSMs) have been extensively studied~\cite{RMP2}. However, the hallmarks of DSMs are still not clear~\cite{kargarian8648,Le8311}. 
Recently, a generalized bulk-boundary correspondence, namely higher-order bulk-hinge correspondence, for DSMs~\cite{wieder2020strong,fang2021dirac,akbar1,zeng2022topological} has been proposed, \ie one-dimensional (1D) higher-order Fermi arcs (HOFAs) are direct and topological consequences of 3D bulk Dirac points. The 3D bulk Dirac points lead to the nontrivial filling anomaly $\eta$~\cite{anomaly1,anomaly2} of the 2D insulating momentum-space plane away from them, which ensures the presence of gapless mid-gap states on 1D hinges. When the 2D plane passes through the bulk Dirac point, the filling anomaly changes from nontrivial to trivial. At last, the gapless mid-gap states form the HOFAs, which connect the hinge projections of the 3D bulk Dirac points. The change of the filling anomaly $\Delta\eta$ is a topological invariant (\ie either a $\mathbb{Z}_2$ or a $\mathbb{Z}_3$ depending on the rotational symmetry), and can be used to characterize the 3D bulk Dirac points~\cite{fang2021dirac}.
Although Ref. \cite{wieder2020strong} proposed some candidates hosting HOFAs, these materials exhibit surface Fermi arcs. Therefore, it is still challenging and highly desirable to find a realistic DSM without surface Fermi arcs but with HOFAs.

In this work, the electronic structures of kagome lattices $\text{Pd}_{3}$Pb$_{2}$X$_{2}$ ($\text{X}=\text{S}$, $\text{Se}$)  are systematically investigated, resulting in the discovery of unique type-I DSMs protected by time-reversal symmetry $\mathcal{\hat{T}}$, inversion symmetry $\mathcal{\hat{I}}$ and three-fold rotational symmetry around $z$-axis $\mathcal{\hat{C}}_{3z}$. 
The absence of surface Fermi arc states in $\text{Pd}_{3}$Pb$_{2}$X$_{2}$ is in contrast with previous well-known DSMs Na$_3$Bi and Cd$_3$As$_2$~\cite{dirac0,dirac2}. However, as a result of the higher-order bulk-hinge correspondence, $\text{Pd}_{3}$Pb$_{2}$X$_{2}$ have a nontrivial topological invariant $\mathbb{Z}_3$ and exhibit symmetry-protected HOFAs and fractional corner charges on the 1D $\mathcal{\hat{C}}_{3z}$-symmetric rods. The hinge projections of bulk Dirac points are connected by the HOFAs, which are hallmarks of $\mathbb{Z}_3$ DSMs. Moreover, strain effect has been investigated, showing that $\text{Pd}_{3}$Pb$_{2}$X$_{2}$ are located at the phase boundaries of three different topological states, \ie type-I DSMs, type-II DSMs, and 3D weak topological insulators (TIs).

%\subsection{Band structures of $\text{Pd}_{3}$Pb$_{2}$S$_{2}$}
%\label{s4def}
%To study the electronic structure information of kagome materials $\text{Pd}_{3}$Pb$_{2}$S$_{2}$, we systematically studied their band structures with and without SOC along the high symmetry $k$ points indicated in Fig. \ref{fig1}(b). 

\begin{figure*}[t]
\includegraphics[width=7.in]{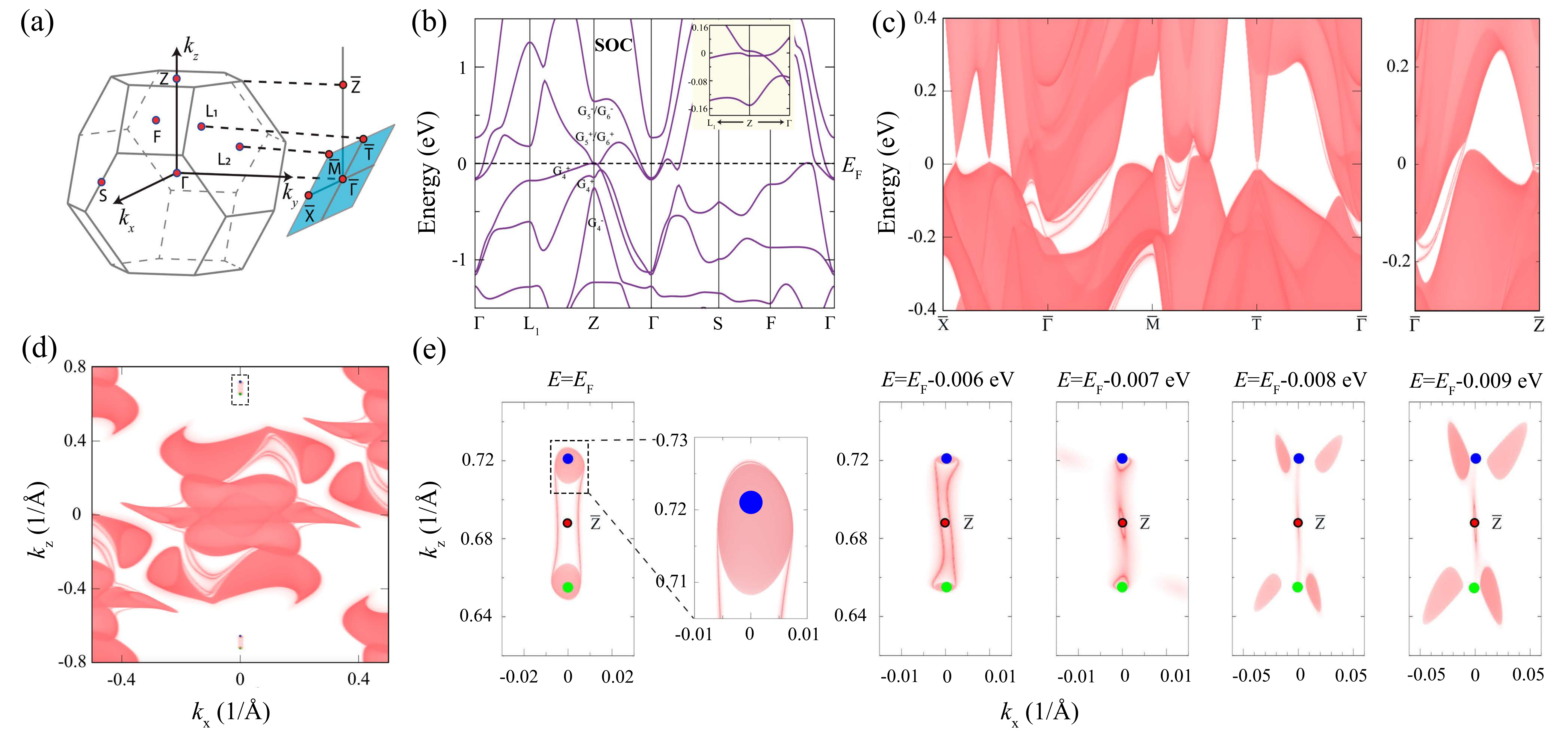}
      \caption{ (color online). Bulk and surface electronic structure of $\text{Pd}_{3}$Pb$_{2}$S$_{2}$. (a) Bulk BZ and $xoz$ surface BZ with high-symmetry points indicated. (b) Bulk band structure along the high-symmetry $k$ points in (a) with SOC included. The insert shows the zoom-in band structure around the Dirac point. 
(c) The local density of states on $xoz$ surface. (d) The $xoz$-surface energy contour 
      with the chemical potential at the Fermi level $\emph{\text{E}}_\text{F}$. The black rectangle indicates the bulk Dirac point projections near the $\bar{\text{Z}}$ point. 
      (e) The zoom-in plots of surface energy contour with different chemical potentials near the bulk Dirac point projections. The energy of the bulk Dirac point is equal to $\emph{E}_{\text{F}}-0.008$ eV. The blue and green dots represent the projected Dirac points. The $\bar{\text{Z}}$ point with coordinate $(k_x, k_y, k_z)=(0, 0, 0.6878)$ in units of $1/\text{\AA}$ is indicated. The $xoz$-surface states in (c)-(e) are calculated by the maximally-localized Wannier function method.}
\label{fig1}
\end{figure*}

$\text{Pd}_{3}$Pb$_{2}$Se$_{2}$ has nearly the same properties as those of $\text{Pd}_{3}$Pb$_{2}$S$_{2}$, so we 
choose $\text{Pd}_{3}$Pb$_{2}$S$_{2}$ as an example for the following discussion.
Fig. \ref{fig1}(b) shows the band structure of $\text{Pd}_{3}$Pb$_{2}$S$_{2}$ with spin-orbit coupling (SOC). 
Due to the strong SOC of the Pb-$p$ orbitals, there is a band inversion between the $\text{G}_5^+/\text{G}_6^+$ bands and the $\text{G}_4^-$ bands labeled by the irreducible representations of the point group $\text{D}_{3\text{d}}$ at the Z point, leading to a type-I Dirac point in the $\Gamma$-Z line, as shown in the inset of Fig. \ref{fig1}(b). 
The Dirac point is different from that of previous known DSMs Na$_3$Bi and Cd$_3$As$_2$, as will be shown below.
Because of $\mathcal{\hat{I}}$, 
the possible topological properties in $\text{Pd}_{3}$Pb$_{2}$S$_{2}$ can be checked by calculating the symmetry-based indicators $Z_{2,2,2,4}=(z_{2,1},z_{2,2},z_{2,3};z_4)$ using the parity-based method~\cite{fu2007z2}. 
According to the parities of the valence states at eight time-reversal-invariant momenta, $Z_{2,2,2,4}$ is computed to be (111;0), guaranteeing the presence of nontrivial topological properties in $\text{Pd}_{3}$Pb$_{2}$S$_{2}$.

%\subsection{Surface states of $\text{Pd}_{3}$Pb$_{2}$S$_{2}$ }
%To clearly show the unique properties of $\text{Pd}_{3}$Pb$_{2}$S$_{2}$, the maximally localized Wannier functions (MLWFs) for $d$ orbitals of Pd, $p$ orbitals of Pb, and $p$ orbitals of S are constructed and they are used to build the Green’s functions of the semi-infinite slabs by using an iterative method (see details in Supplemental Material). 
%we chose the major obitals  $d$ orbitals of Pd, $p$ orbitals of Pb, and $p$ orbitals of S to construct the maximally localized Wannier functions (MLWFs) 
To clearly show the unique properties of $\text{Pd}_{3}$Pb$_{2}$S$_{2}$, the local density of states on $xoz$ surface are calculated by using Green's function method~\cite{rmp}, as shown in Fig. \ref{fig1}(c). It is easy to find that both the $\bar{\Gamma}$ point and the ${\bar{\text{T}}}$ point have a surface Dirac cone at around -0.2 eV and -0.008 eV, respectively. It is worth noting that at the $\bar{\Gamma}$ point only the upper Dirac cone could be clearly seen, while the lower Dirac cone is buried into the bulk valence states. 
The existence of these two Dirac cones in the first Brillouin zone (BZ) of $xoz$ surface is consistent with the calculated nontrivial $Z_{2,2,2,4}$.
Next, the surface states around the projections of the bulk Dirac points are carefully studied. 
In the $\bar{\Gamma}-\bar{\text{Z}}$ line, the bulk Dirac point results in a clear bulk band touching around the Fermi level (Fig. \ref{fig1}(c)). In addition, a surface Dirac cone exists at the $\bar{\text{Z}}$ point because the $\bar{\text{Z}}$ point is equal to the $\bar{\text{T}}$ point. 
To check if there are surface Fermi arcs in $\text{Pd}_{3}$Pb$_{2}$S$_{2}$, the $xoz$-surface energy contours are calculated, as shown in Fig. \ref{fig1}(d) and (e). It seems that there are two arc-like surface states. However, after careful examination, they are found not to merge into the bulk electron pockets, but to surround them, \ie they are continuous in momentum space and not Fermi arc states. When the energy of the bulk Dirac point is chosen as the chemical potential, it is clear that there is no surface Fermi arc states connecting the two projected Dirac points (Fig. \ref{fig1}(e)), which is different from the previous well-known DSMs Na$_3$Bi and Cd$_3$As$_2$.
To get a quantitative insight of the Dirac points, an effective low-energy $4\times4$ $k\cdot p$ model is constructed for the two inverted doubly-degenerate bands around the $\text{Z}$ point with irreducible representations $\text{G}_5^+/\text{G}_6^+$ and $\text{G}_4^-$, respectively.  It is found that the antidiagonal terms for $\text{Pd}_3\text{Pb}_2\text{S}_2$ have significantly large coefficients and cannot be ignored due to its specific structure, which is a main reason for the absence of surface Fermi arcs.

%\subsection{Filling anomaly, HOFAs and fractional corner charges}

\begin{figure}[t]
\includegraphics[width=3.4in]{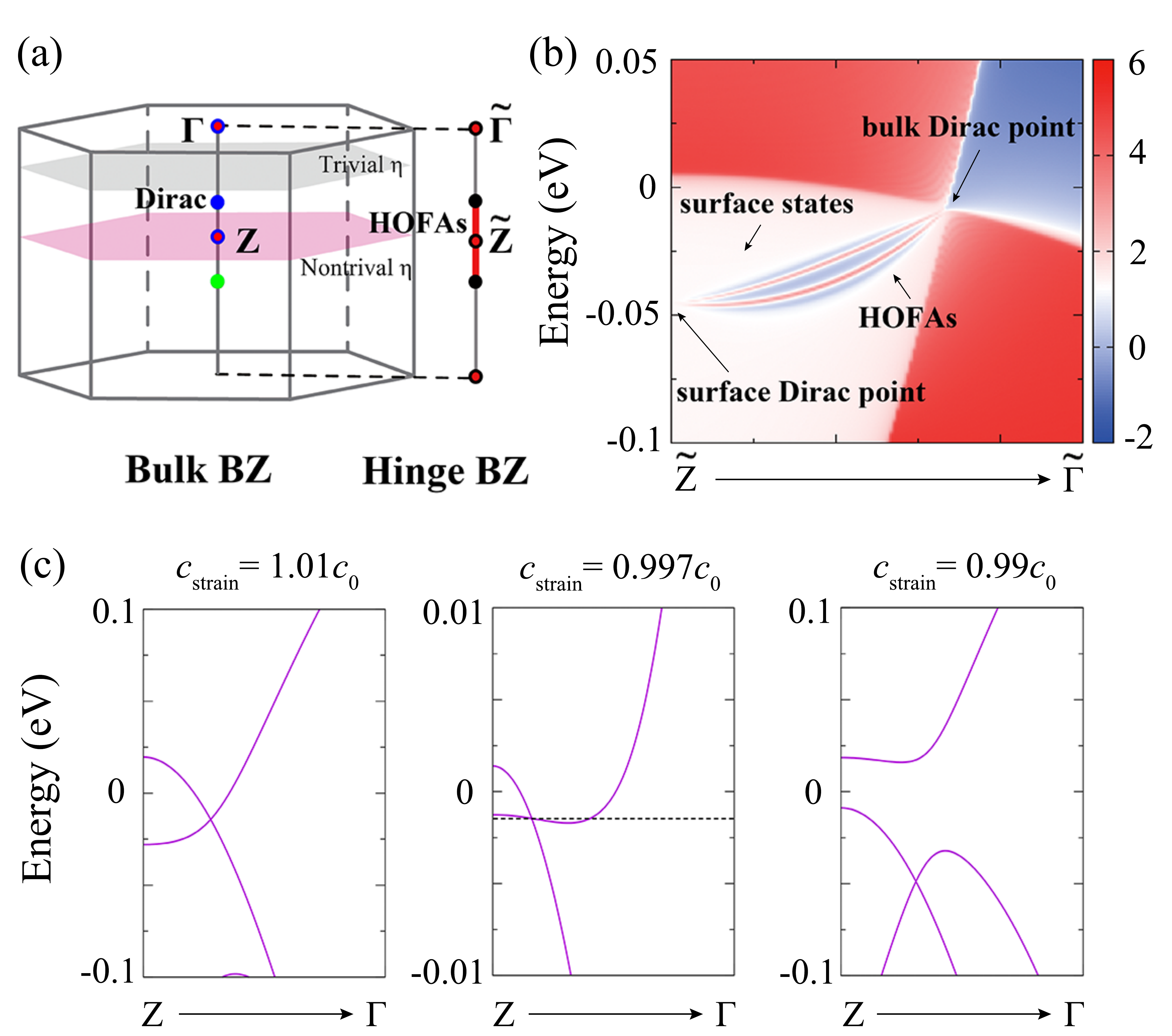}
      \caption{(color online). The HOFAs in $\text{Pd}_{3}$Pb$_{2}$S$_{2}$ and band structures of strained $\text{Pd}_{3}$Pb$_{2}$S$_{2}$.
      (a) Schematic of the HOFAs. (b) Band structure of $\text{Pd}_{3}$Pb$_{2}$S$_{2}$ projected onto its 1D [001] hinge. The $\Tilde{\text{Z}}$ point is chosen as the origin point. The hinge spectrum is calculated based on the tight-binding model of an A-A stacked triangle lattice. (c) The band structures of strained $\text{Pd}_{3}$Pb$_{2}$S$_{2}$. $c_0$ is the experimental lattice parameter.}
\label{fig2}
\end{figure}

For each 2D $\mathcal{\hat{C}}_{3z}$-invariant insulating slice in momentum space, the filling anomaly is well defined, and related to the corner charge $\mathcal{Q}_c$ of the specific [001] hinge as~\cite{anomaly1}
\begin{eqnarray}
\mathcal{Q}_c=\frac{\eta}{3}|e| ~~\text{mod}~2,
\label{kpsoc} 
\end{eqnarray}
where \emph{e} denotes the elementary charge. 
For the $k_z$ plane with $-k_\text{D}<k_z-\text{Z}<k_\text{D}$ ($k_\text{D}$ plane indicates the plane with the Dirac point), the nontrivial $\eta$ thus guarantees the presence of fractional corner charges on 1D hinges, which finally form HOFAs, as shown in Fig. \ref{fig2}(a). However, $\eta$ is trivial for the $k_z$ plane with $k_z-\text{Z}>k_\text{D}$ or $k_z-\text{Z}<-k_\text{D}$. Therefore, although the surface states near the surface projections of the 3D bulk Dirac points in $\text{Pd}_{3}$Pb$_{2}$S$_{2}$ are not topologically protected, their hinge projections are the terminations of the topological HOFAs, which are hallmarks of DSMs. For the system with Dirac points protected by time-reversal symmetry, inversion symmetry and three-fold rotational symmetry, the change of the filling anomaly $\Delta\eta$ is an even number and defined mod 6. Therefore, $\Delta\eta$ features a $\mathbb{Z}_3$ group structure and is a topological invariant for characterizing the DSMs.
For $\text{Pd}_{3}$Pb$_{2}$S$_{2}$, the filling anomaly of the plane with $-k_\text{D}<k_z-\text{Z}<k_\text{D}$ is computed to be 4, while it is 0 for the plane with $k_z-\text{Z}>k_\text{D}$ or $k_z-\text{Z}<-k_\text{D}$. Therefore, the invariant $\Delta\eta$ is equal to 4, signaling the existence of HOFAs on 1D hinges of $\text{Pd}_{3}$Pb$_{2}$S$_{2}$, as shown in Fig. \ref{fig2}(b). It is found that the hinge projection of the surface Dirac cone at the $\bar{\text{Z}}$ point exists at the $\Tilde{\text{Z}}$ point, while the HOFAs connect the hinge projections of the bulk 3D Dirac cones to hinge projection of the surface Dirac cone at the $\Tilde{\text{Z}}$ point. The results are consistent with the calculated filling anomaly.

%\subsection{Phase transitions in $\text{Pd}_{3}$Pb$_{2}$S$_{2}$ }
We also find the type-I DSM can be driven into type-II DSM or 3D weak TI by uniaxial strain effect along the $z$-axis, as shown in Fig. \ref{fig2}(c). 
The calculations show that the expansion strain is beneficial to the formation of the type-I Dirac points (\ie $c_{\text{strain}}=1.01 c_0$). 
When a small compressive strain is applied (\ie $c_{\text{strain}}=0.997 c_0$), the type-I Dirac point evolves into a type-II Dirac point. Interestingly, when the compressive strain is further increased (\ie $c_{\text{strain}}=0.99 c_0$), the compound changes from the type-II DSM state to an insulating state due to the band order change of the $\text{G}_5^+/\text{G}_6^+$ bands and $\text{G}_4^+$ bands. 
As these states have the same parity, the symmetry-based indicators $Z_{2,2,2,4}$ of the insulating state is still (111;0), but actually corresponds to a 3D weak TI. Therefore, our results provide feasible platforms for the study of these unique DSMs and the related phase transitions.

%\section{Discussion and Conclusion}

Being able to interact with other quantum states is one of the charming advantages of topological states. 
For example, the interplay between 3D Dirac points and the superconductivity state may lead to 3D Dirac superconductor, featuring gapless Dirac points in the 3D bulk spectrum and Majorana Fermi-arc surface states~\cite{cui2020type}. Moreover, when both superconductivity and magnetic field are considered, the 3D weak TIs pave a way to achieve Majorana modes, which can be used to realize fault-tolerant quantum computing. Recently, the superconductivity has been observed in $\text{Pd}_{3}$Pb$_{2}$Se$_{2}$ after high pressure is applied~\cite{yu2020pressure}. Therefore, $\text{Pd}_{3}$Pb$_{2}$X$_{2}$ are fertile ground for studying the interactions of DSMs (or TIs) with other states.

In conclusion, we propose that the kagome lattice $\text{Pd}_{3}$Pb$_{2}$X$_{2}$ are type-I DSMs characterized by nontrivial $\mathbb{Z}_3$. Compared with Na$_3$Bi and Cd$_3$As$_2$, $\text{Pd}_{3}$Pb$_{2}$X$_{2}$ feature closed Fermi surface states, illustrating that the presence of Fermi arc surface states is not a necessary condition for DSM. As direct consequences of the Dirac points in $\text{Pd}_{3}$Pb$_{2}$X$_{2}$, there are fractional corner charges, and HOFAs on its specific hinges. The HOFAs connect the hinge projections of bulk Dirac points. In addition, the type-I DSMs $\text{Pd}_{3}$Pb$_{2}$X$_{2}$ can be driven into type-II DSMs or 3D weak TIs by external strain. 
\ \\

\noindent \textbf{Conflict of interest}

The authors declare that they have no conflict of interest.
\ \\

\noindent \textbf{Acknowledgments}

We thank Xianxin Wu for useful discussions. This work was supported by the National Natural Science Foundation of China (Nos. 11974076, 11925408, 11921004, and 12188101), the Key Project of Natural Science Foundation of Fujian Province (2021J02012), the Ministry of Science and Technology of China (2018YFA0305700), the Chinese Academy of Sciences (Nos. XDB33000000 and CAS-WX2021SF-0102), the K. C. Wong Education Foundation (GJTD-2018-01), the Key Research Project of Zhejiang Lab (No. 2021PB0AC01). Changming Yue was supported by the Swiss National Science Foundation (200021-196966).

\ \\
\noindent \textbf{Author contributions }

Hongming Weng, Wei Zhang and Simin Nie proposed and supervised the project. Jia Chen, Simin Nie and Danwen Yuan carried out DFT calculations. Simin Nie, Congcong Le, Zhijun Wang, Changming Yue and Wei Zhang did the theoretical analysis. All authors contributed to writing the manuscript.

\noindent \textbf{Appendix A. Supplementary materials }

Supplemental Materials to this article can be found online at XXX.
%\bibliographystyle{unsrt}
%\bibliography{pdpbs}

\end{document}